# Three-dimensional modeling of chiral nematic texture evolution under electric switching

Vianney Gimenez-Pinto[*]|| Sajedeh Afghah, and Robin L. B. Selinger[+]

*Advanced Materials and Liquid Crystal Institute, Kent State University, Kent, OH 44242*

*||Corresponding author: gimenez-pintov@lincolnu.edu*

*+Corresponding author: rselinge@kent.edu*

*\*Current Address: Department of Science, Technology and Mathematics, Lincoln University, Jefferson City, MO 65101*

**Abstract**: Chiral nematic liquid crystals exhibit both a helical planar ground state with uniform twist and a metastable defect-rich focal conic texture, and can be switched between the two microstructures via application of transient voltage pulses. In this work, we model these electrically-induced texture transitions using finite difference methods to examine resulting microstructural evolution, the first time this transition has been modeled in three dimensions. We analyze the planar to focal conic, focal conic to planar, and planar to planar transitions depending on voltage pulse magnitude. We consider first the special case of chiral nematics with matched twist and bend elastic constants. Results show a variety of defect-rich morphologies in the disordered focal conic texture and demonstrate a fast recovery of the planar ground state on switching without formation of a transient planar state. We evaluate both texture microstructural evolution as well as cell capacitance. Beyond the single elastic constant approximation, we evaluate the planar to transient-planar as well as the planar to Helfrich-deformed transitions in simulations of a liquid crystal compound with different elastic constants. Our methods represent the evolving microstructure as a uniaxial director field, with relaxation dynamics calculated from a tensor representation so that half charge disclination defects are not suppressed. We discuss potential application of these computationally efficient three-dimensional modeling approaches for design and optimization of chiral nematic devices.

**Introduction**

Liquid crystals' inherent dielectric anisotropy enables control of their molecular orientation by applying an external electric field. In a rich assortment of applications, liquid crystals are confined by surfaces that impose anchoring forces on molecular orientation, while internal stresses also arise from the material's elastic response. Competition among these effects determines the resulting microstructural evolution of the liquid crystal's director field and resulting change of optical properties under an applied electric field. While this mechanism serves as a common foundation for conventional liquid crystal displays and optical devices, the distinctive response of these materials to external field can serve as the basis for a wide range of other technologies. Examples include polymer-dispersed liquid crystals (PDLC) for smart window applications [1] and polymer-stabilized cholesteric displays (PSCD) [2][3]. These hybrid devices are liquid-crystal/polymer composites in which geometry, confinement and anchoring conditions with the surrounding polymer produce a rich variety of mesophase textures. These switchable textures create states that produce scattering, transmission, and – in the case of chiral nematic liquid crystals – selective reflection of light with a characteristic wavelength. Beyond well-known display technology, recent experimental studies also report the potential development of micro-lenses based on specific confined geometries adopted by chiral nematic films submerged in water [4]. Furthermore, confinement of chiral nematic liquid crystals in micro-channels with different aspect ratios can produce either striped textures or an array of bubble defects [5]. Depending on geometry and anchoring conditions, chiral nematics confined in a spherical volume can exhibit complex textures including topological skyrmions stabilized by strong homeotropic anchoring [6] and defect lines varying from extended configurations to perfect three-fold knots [7]. There is also growing interest in the study of N* textures in spherical confinement, where the switching mechanism can be given by variable anchoring strengths [8] and/or an active helical pitch [9], while geometries can vary from thick hollow spheres to thin shells [10].

Cholesterics' ground state is a helical planar texture with a uniform twisted director field characterized by the pitch $p$, the length over which the director twists by 360°. Cholesterics also exhibit a metastable defect-rich focal conic texture that scatters light in all directions. Because both planar and focal conic textures are stable at room temperature and have sharply contrasting optical properties, these materials form the basis for information display devices that can maintain an image without the use of electric power after switching [2].

Recent experimental studies of cholesteric liquid crystals with matched twist and bend elastic constants [11] have demonstrated a fast-switching mechanism between planar and focal conic textures. Under a high enough voltage pulse, such materials relax directly to a defect-free planar texture with uniformly oriented helical structure. If elastic constants are not matched, switching is significantly slower because the material first forms a transient planar state with pitch $p'=(K_{33}/K_{22})\,p$, where $K_{33}$ is the bend elastic constant and $K_{22}$ is the twist elastic constant. The planar ground state eventually forms via nucleation and growth from the transient planar state, often creating oily streak defects.

To examine the switching mechanism proposed in [8], first we perform simulation studies of this bi-stable switching mechanism in chiral nematic liquid crystals in the special case of matched elastic constants, comparing their relaxation behavior after applied pulses of different voltages. We show *for the first time* a three-dimensional simulation study of the planar to focal conic; focal conic to planar; and planar to planar transitions as a function of voltage pulse amplitude. Next, we study the slower switching behavior in chiral nematic materials with different elastic constants. Simulation shows the expected formation of the transient planar texture with pitch defined by the ratio of bend and twist elastic constants. Also the well-known Helfrich undulations are observed.

Our findings demonstrate and confirm the mechanism proposed to explain experimental observations in [11]: when the applied voltage exceeds a threshold, the focal conic relaxes directly from an untwisted state to the helical planar ground state. Simulations demonstrate the absence of the transient planar state in the case of matched elastic constants and resulting fast relaxation to the defect-free planar ground state. In addition, we observe the emergence of disordered focal conic textures with a rich variety of defects morphologies and undulations in the helical axis along a wide range of voltages. We note that no topological characterization has been performed in disordered chiral textures stabilized by weak planar anchoring, thus the importance of reporting these arising morphologies in a fully 3-D model. Given that a three-dimensional model allows the formation of textures that cannot be observed in a two-dimensional approach, revisiting these textures transitions becomes necessary. Our results are in qualitative agreement with observed bistable switching behavior in chiral nematics. Our simulations show nucleation and growth of defect-free "perfect planar" textures from the anchoring surfaces, while disordered and undulated states appear due to defects that nucleate in the bulk after switching or by the addition of thermal noise.

Remarkably, our 3-D study on texture switching offers the first simulation study of the different morphologies and spatial variations that helical axis exhibits in the focal conic texture, including the emergence and sequence of rounded, elongated and helix-undulated morphologies that arise during the focal conic to planar transition.

**Modeling details and assumptions**

We implement a finite difference approach for modeling chiral liquid crystal textures including their co-evolving electric field response. Due to nematics' inversion symmetry, it is necessary to implement a free energy representation based on the $Q_{ij}$ nematic order parameter tensor as documented in the literature [1][12] in order to avoid suppressing half-charge disclination defects. This Frank free energy may be written as

$$f = \frac{1}{12}(-K_{11} + 3K_{22} + K_{33})G_1 + \frac{1}{2}(K_{11} - K_{22})G_2 + \frac{1}{4}(-K_{11} + K_{33})G_6 - q_0 K_{22} G_4 \\ + \frac{1}{2}(K_{22} + K_{24})G_5 \quad (1)$$

where $K_{11}$, $K_{22}$ and $K_{33}$ are the elastic constants corresponding with splay, twist and bend deformations of the nematic director and the scalars $G_i$ are

$$G_1 = Q_{jk,l}Q_{jk,l} = 2\left[\left(\nabla \cdot \vec{n}\right)^2 + \left(\vec{n} \cdot \nabla \times \vec{n}\right)^2 + \left(\vec{n} \times \nabla \times \vec{n}\right)^2 - \vec{\nabla} \cdot \left(\vec{n}\vec{\nabla} \cdot \vec{n} + \vec{n} \times \vec{\nabla} \times \vec{n}\right)\right]$$

$$G_2 = Q_{jk,k}Q_{jl,l} = \left(\nabla \cdot \vec{n}\right)^2 + \left(\vec{n} \times \nabla \times \vec{n}\right)^2$$

$$G_4 = e_{jkl}Q_{jm}Q_{km,l} = -\vec{n} \cdot \nabla \times \vec{n} \qquad (2)$$

$$G_5 = G_2 + G_4^{\ 2} - \frac{1}{2}G_1 = \vec{\nabla} \cdot \left(\vec{n}\vec{\nabla} \cdot \vec{n} + \vec{n} \times \vec{\nabla} \times \vec{n}\right)$$

$$G_6 = Q_{jk}Q_{lm,j}Q_{lm,k} = 2\left(\vec{n} \times \nabla \times \vec{n}\right)^2 - \frac{1}{3}G_1$$

Solving Eq. 1 and 2 is computationally intensive when applied to large three-dimensional systems, particularly when combined with a simultaneous solution of the material's dielectric response to a time-varying applied voltage. To model time evolution of the director field, *we implement relaxation methods based on a simplified free energy functional derived from the $Q_{ij}$ tensor formalism*. Our approach has the distinctive advantage that it is consistent with the inversion symmetry of nematics and can be used for computation of large three-dimensional systems. In the aim to develop an efficient modeling method[+] for devices at the laboratory scale, spatial discretization is large compared to the defect-core size (~10 nm). Thus, we do not solve for variations of the scalar order parameter *S* within defect-cores, and reduce computational complexity in the problem by assuming a spatially uniform scalar order parameter. Likewise, as a simplifying approximation, we neglect biaxiality in molecular ordering. These approximations reduce the number of degrees of freedom for the symmetric, traceless $Q_{ij}$ tensor from five to two at each point in space, thus allowing nematic ordering to be described simply by director-explicit terms.

*Single elastic constant approach:* To model the liquid crystal bulk, we define a cubic lattice with nearest neighbor spacing *c* and use the lattice's symmetries to simplify the $Q_{ij}$ tensor free energy functional. Assuming a single elastic constant approximation and neglecting the surface integral term $G_5$, we use a finite difference technique to re-express $G_1$ and $G_4$ as a function of cross and dot products of the director on nearest neighbor sites – See Supplementary Information – and obtain a free energy with the form:

$$F = \frac{K}{2} c \sum_{<a,b>} \left(1 - \left(\vec{n}^a \cdot \vec{n}^b\right)^2\right) - K q_0 c \sum_{<a,b>} \left(\vec{n}^a \cdot \vec{n}^b\right)\left(\vec{n}^a \times \vec{n}^b \cdot \vec{r}^{ab}\right) \quad (3)$$

This free energy functional is explicitly independent of director sign, and allows topological defects with fractional charge $m = -½, + ½$. Like the Lebwohl-Lasher model, this approach includes the single elastic constant approximation and lattice discretization. We calculate time evolution of the director field by solving for director rotation dynamics in the overdamped limit, rather than by a Monte Carlo approach. Similar studies on Lebwohl-Lasher Molecular Dynamics have been implemented for investigating the nematic-isotropic phase transition [13][14].

The torque $\tau$ acting on the director located at the lattice site $a$ follows $\tau_l = -e_{lki} n_k^a \frac{\partial F}{\partial n_i^a}$. As implemented here, the model only allows director rotation, neglecting backflow effects. Local director at each site updates according to $\vec{n}_t = \vec{n}_{t-1} + \left(\vec{\omega} \times \vec{n}_{t-1}\right)\Delta t$, with angular velocity following $\vec{\omega} = \xi \vec{\tau}$ and damping factor $\xi \propto \gamma_1$. Next, we renormalize the director at each lattice site. The elastic constants, rotational viscosity $\gamma_i$ and dielectric constant parameters $\varepsilon_i$ are set to values corresponding with a commonly used commercial liquid crystal compound: $K = K_{22} = 13.0$ x $10^{-12}$ N, $\gamma_1 = 0.3875$ Pa s, $\varepsilon_{//} = 22.0$ and $\varepsilon_\perp = 5.2$. Simulation lattice spacing was set to $c = 0.333$ μm and system size corresponds with a cube of side 10.0 μm. Cholesteric liquid crystal has pitch $p = 3.33$ μm; so $q_0 = 6\pi/10$ μm$^{-1}$. We note that modeling a different liquid crystal compound will affect the textures reported in this work. These textures are deeply ingrained with the inherent elastic constants of the material, arising from the mesogens' molecular structure.

*Distinct elastic constants approach*: In order to simplify the implementation of different elastic constants in the $Q_{ij}$ tensor free energy functional (Eqs. 1 and 2), we consider a cubic lattice and apply the relationships: $\frac{\delta F}{\delta n_i} = \frac{\delta F}{\delta G_j} \frac{\delta G_j}{\delta n_i}$ ; $\frac{\delta G}{\delta n_i} = \frac{\delta G}{\delta Q_{rm}} \frac{\delta Q_{rm}}{\delta n_i} = \left(n_r \delta_{ni} + n_n \delta_{ri}\right) \frac{\delta G}{\delta Q_{rm}} = 2 n_r \frac{\delta G}{\delta Q_{ri}}$ ;

$\frac{\delta G}{\delta Q_{ri}} = \frac{\partial G}{\partial Q_{ri}} - \frac{\partial}{\partial x_n}\left(\frac{\partial G}{\partial Q_{ri,n}}\right)$; and $Q_{ri,n} = \frac{\delta Q_{ri}}{\delta x_n}$. Then, the director evolution can be re-written as:

$$\Delta n_i^{t+1} = \alpha \left\{ \Delta x^2 \left[ -\frac{1}{12}\left(-K_{11} + 3K_{22} + K_{33}\right) H_1^t(i) - \frac{1}{2}\left(K_{11} - K_{22}\right) H_2^t(i) \right.\right.$$

$$\left.\left. -\frac{1}{4}\left(-K_{11} + K_{33}\right) H_6^t(i) - q_0 K_{22} H_4^t(i) \right] - \Delta x \left[ -\frac{1}{2} K_{24} H_5^t(i) \right] \right\}$$

where $H_m^t(i) = \frac{\partial G_m}{\partial n_i}$. Director components are normalized each time they are updated. This representation avoids the complication of calculating the interaction between two neighboring directors that are anti-parallel. Further details on this method can be found in [5]. For computational efficiency, the director was GPU-accelerated via CUDA implementation, allowing full computations on a system of the same size as the single elastic constant approach, below the one-hour mark. Given the existence of different values for the splay $K_{11}$, twist $K_{22}$, bend $K_{33}$ and saddle-splay $K_{24}$ elastic constants, this formulation allows the formation of the transient planar texture as well as Helfrich undulations in the chiral nematic system. Unless stated otherwise, simulation parameters in the different elastic constant study are: $K_{11}$ = 6.01 pN, $K_{22}$ = 3.82 pN, $K_{33}$ = 8.56 pN, $K_{24}$ = 5.97 pN and so $K_{33}/K_{22}$ = 2.24. The director relaxation factor is 0.002. Further details can be found in [15].

*Applied electric field:* An external electrical field is applied by imposing a voltage pulse between the two surfaces in the x-direction. Liquid crystal devices are normally switched using an AC field to prevent charged ionic species from aggregating at the electrodes. For simplicity, ion motion effects are not included in the model. We calculate response to a DC field, and use material dielectric properties matched to the AC frequency of interest. However, the effect of frequency in the switching behavior can be modeled via direct inputting the frequency dependence of the liquid crystal dielectric constants. Further work can also extend the model to include backflow effects. More details in the AC and DC dielectric response of liquid crystals have been reported in experimental studies [16][17].

The electrical contribution to the free energy is $F_{electric} = -\vec{D} \cdot \vec{E}$ where $\vec{D} = \varepsilon_0 \varepsilon_{ij} \vec{E}$ is the electric displacement vector, and $\varepsilon_{ij}$ is the anisotropic dielectric tensor for the liquid crystal [18][19]. The local dielectric tensor in the liquid crystalline bulk is $\varepsilon_{ij} = \varepsilon_\perp + \left(\varepsilon_\parallel - \varepsilon_\perp\right) n_i n_j$ [1], which varies as a function of position and time as the director field relaxes. Thus, we have a dielectric tensor that depends on the nematic director; while at the same time contributes to the electric torque affecting nematic orientation.

We calculate the electrostatic potential, electric field and dielectric tensor by solving $\nabla \cdot \vec{D} = 0$ at the beginning of each time step, via over-relaxation [20] with convergence criterion $\langle \Delta \varphi \rangle < 1 \times 10^{-5}$. Then, we apply electrical torque and update the nematic director [19]; the dielectric tensor at each point in the system is thus changed. At the next time step, we use the calculated voltage distribution V(x,y,z) from the previous step as a first guess and use over-relaxation to again solve $\nabla \cdot \vec{D} = 0$. To monitor texture evolution, we calculate the cell capacitance *C* at each time step using the induced surface charge on both substrates under voltage. After the voltage pulse, we measure *C* with a test voltage $V_1 = 1V$, small enough that it does not significantly affect the final cholesteric texture.

Surface anchoring terms are added to the free energy to describe interaction of the liquid crystal with the confining substrates. The free energy term describing anisotropic interfacial energy is $\vartheta_s = \int f_s^- ds^- + \int f_s^+ ds^+$. Interfacial energy density follows $f_s = W_\varsigma (\vec{n} \cdot \vec{\varsigma})^2 + W_\eta (\vec{n} \cdot \vec{\eta})^2$ where $(\vec{\varsigma}, \vec{\eta}, \vec{\iota})$ is an orthonormal triplet and $\vec{\iota}$ is the easy axis of orientation [21]. Unidirectional anchoring has easy axis $\vec{\iota} = \vec{y}$, thus $W_\varsigma$ and $W_\eta$ are the weak azimuthal and polar anchoring strengths respectively ($W_a = W_p = \sim 10^{-6}$ N/m).

Previous work by Anderson, et al. [22] demonstrated the feasibility of modeling chiral nematics at the mesoscopic level in two-dimensions, inspiring our extension to three dimensions for better modeling of real devices. As nucleation and growth behave differently in two vs. three dimensions, a full three-dimensional description is needed.

**Equal Elastic constants: Modeling fast switching planar-to-planar, planar-to-focal conic, and focal conic-to-planar**

Using the simulation approach described above, we modeled the electrical switching of chiral nematic textures in a bulk of liquid crystal material. First, we consider a planar initial state with a defect-free helical texture with the helical axis along the *x*-direction. We apply a voltage pulse of magnitude $V_0$ ranging from 4V - 60V during 50000 time steps, corresponding to 500ms, after which the sample relaxes to an equilibrium texture.

For an applied voltage pulse $V_0 = 10$V, we observe *planar-to-planar* switching behavior, as shown in Fig. 1. In this case, the planar texture relaxes from one of the substrates after $V_0$ is removed. While $V_0$ is sufficient to unwind the helix to a vertical state, anchoring favors a tilt from vertical orientation near the substrates. Thus, anchoring energy drives the planar texture to nucleate and grow from the electrode surface. The total energy and capacitance measurements show that the system is fully relaxed and stable after switching. While the final state maintains the same chirality as the initial state, we note that its pitch is slightly longer, giving the appearance of a phase shift. This final texture pitch arises as a consequence of competition among all elastic (*K*) free energy terms, incorporating also the contribution from anchoring (*W*) energy. On the other hand, the initial state in the simulation was parameterized to match the $q_0$ parameter, which only appears in the chiral term of the free energy functional. For applied voltages smaller than 10V, the field is not strong enough to fully unwind the helix and the planar texture persists after relaxation. Microstructural evolution of this switching can be found in Supplementary Videos 1-3.

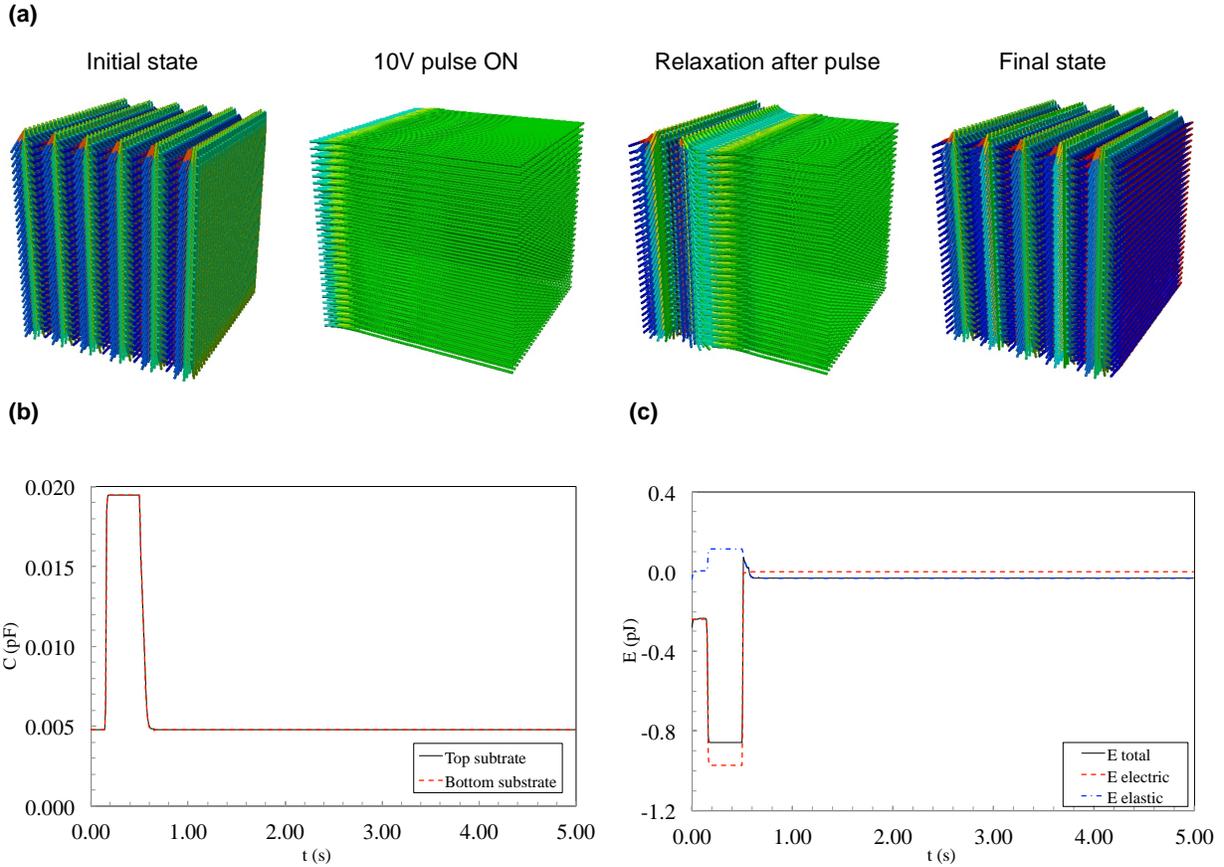

**Figure 1** (Color online) Planar-to-planar transition after a 10V pulse is applied to an initial perfect-planar texture. a) Microstructural evolution of the 3-D nematic director during the planar-to-planar switching. b) Capacitance of the liquid crystal cell. c) Elastic, electric and total energy of the system.

For intermediate applied voltages $V_0$ = 15 to 40V, after the voltage is removed, the helix grows from both sample substrates but it presents defects in the helical axis – see Supplementary Videos 4 - 6. The system finally relaxes to a texture with multiple defects and helical axis oriented at random directions. Figure 2 shows the microstructural evolution of the *planar-to-focal conic* transition with an applied voltage pulse of 15V. We note that these disordered helical textures still show some periodicity while coexisting with defects and undulations of the helix axis.

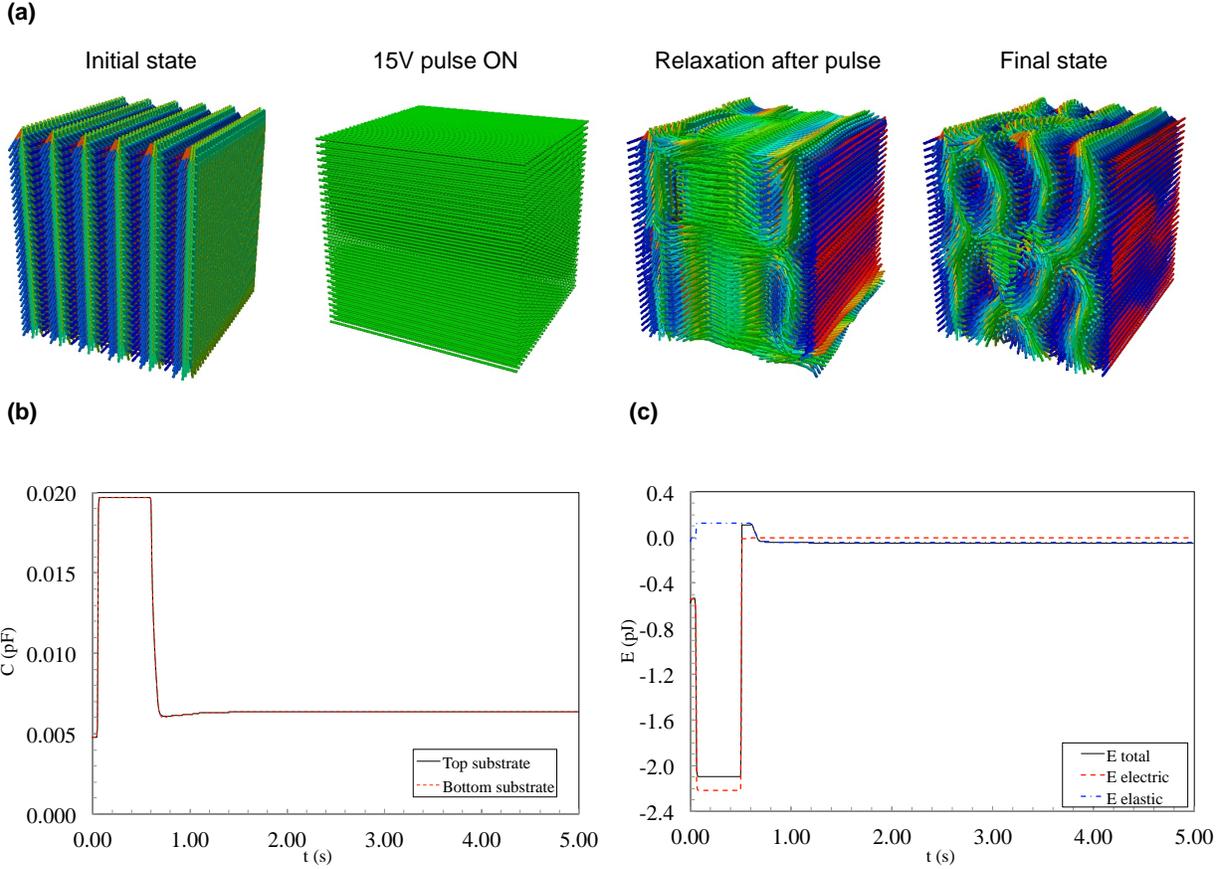

**Figure 2** (Color online) Planar-to-focal conic transition, under a transient voltage pulse of $V_1=15V$ applied to an initial planar texture. a) Microstructural evolution of the 3-D nematic director. b) Capacitance of the liquid crystal cell. c) Elastic, electric and total energy of the system. We observe this planar to focal conic texture transition for voltage pulse of magnitude between 15V and 40V.

With an applied voltage pulse $V_0 = 60V$, the simulation again shows a *planar-to-planar* transition, as shown in Figure 3. The final state is a perfect defect-free planar texture, that is, the helical axis is uniformly oriented perpendicular to the sample substrates and no defects are present (Supplementary Video 7). Similarly to the case in low-voltage planar-to-planar switching, the final state presents a longer pitch than initial state, due to the interplay of anchoring and all elastic terms in *f*. Both states have the same chirality. The observation of these *planar-to-planar* transitions at small and large voltages and *planar-to-focal conic* transitions at intermediate voltages agrees with the bistable switching behavior of cholesteric liquid crystal textures widely reported in the literature [1][12].

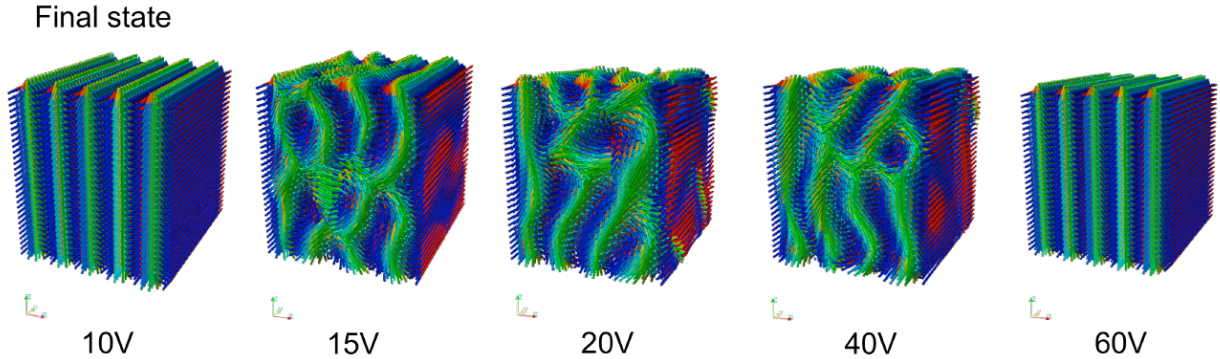

**Figure 3** (Color online) Final state after a transient voltage pulse is applied to a cholesteric liquid crystal initially in the planar state. Lowest and highest magnitude voltage pulses both produce a planar-to-planar transition, while a range of intermediate magnitude voltage pulses produce a planar-to-focal conic texture transition.

Next, we model *focal conic-to-planar* transitions. We take as our initial state the defect-rich texture obtained after full relaxation of the planar-to-focal conic switching under a voltage pulse of 15V, as shown in Figure 3. Figure 4 shows the *focal conic-to-planar* transition with an applied voltage pulse of 60V. This strong electric field produces a complete helical unwinding in the texture, effectively erasing the defect-rich focal conic state and allowing the system to relax back to the defect-free planar state. By contrast, small and intermediate applied voltages are insufficient to erase disorder in the initial focal conic state. Snapshots of the relaxed textures after electrical switching with voltage pulses from 4-60V are shown in Figure 5. For pulses ≤ 40V, the final disordered textures exhibit defects in the helix periodicity as well as undulations in the overall helical axis of the texture. Observed defect structures vary with the strength of the applied voltage (see Supplementary Videos 8 – 17). When $V_0 \leq 12$V, the defect textures exhibit different enclosed morphologies: for a voltage of 4V helical periodicity shows a rounded circle-like morphology while the 10V texture presents elongated enclosed shapes in its periodicity. For 15V $< V_0 <$ 40V, fully enclosed shapes are no longer seen. Disordered textures show up as undulated stripes with a few defects in the helix periodicity.

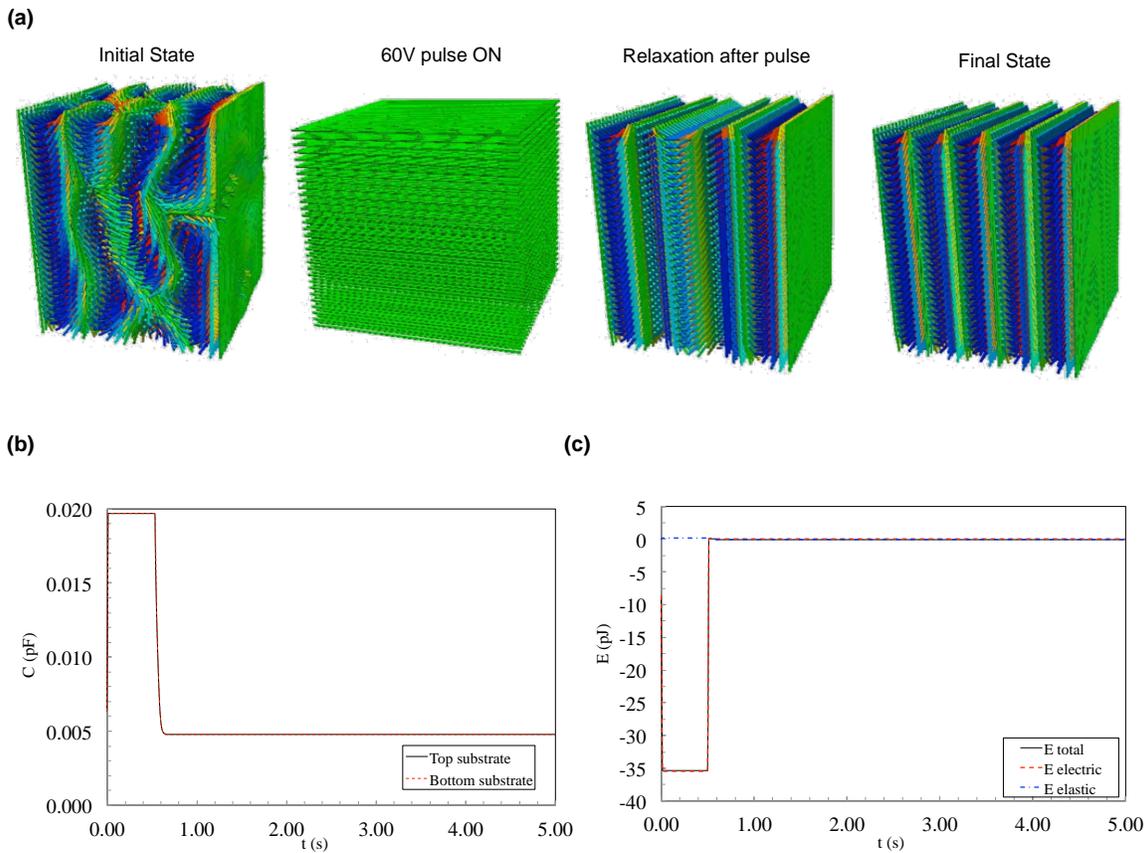

**Figure 4** (Color online) Disordered helical-to-planar transition when a voltage pulse of 60V is applied to the disordered helical texture. a) Microstructural evolution of the 3-D nematic director. b) Capacitance of the liquid crystal cell. c) Elastic, electric and total energy of the system.

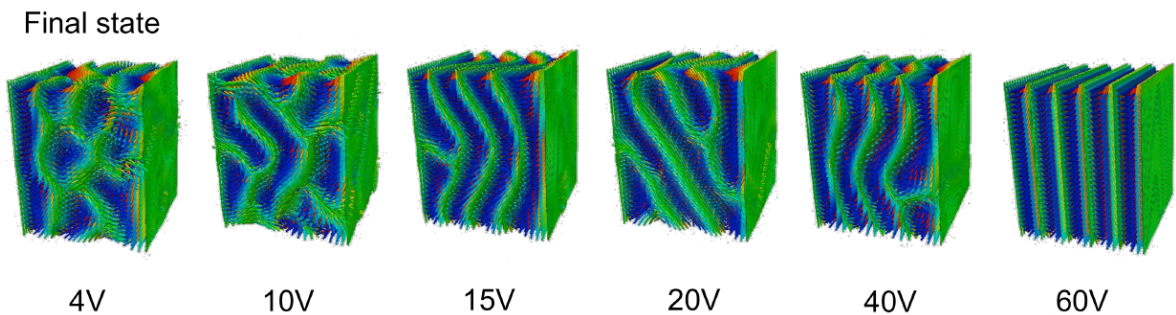

**Figure 5** (Color online) Summary of disordered helical (focal conic) - planar texture sequence relaxed after voltage pulse is applied to a focal conic texture in simulations with weak anchoring conditions.

While the bistable switching behavior in chiral nematics has been widely reported in the literature [1][12], this model presents for the first time a 3-D simulation of the emerging perfect-planar and defect-rich focal conic textures at different voltages from both planar and disordered helical states. These simulations are able to capture different morphologies – from rounded and elongated enclosed morphologies to undulated stripes – in the periodicity of the disordered focal conic states.

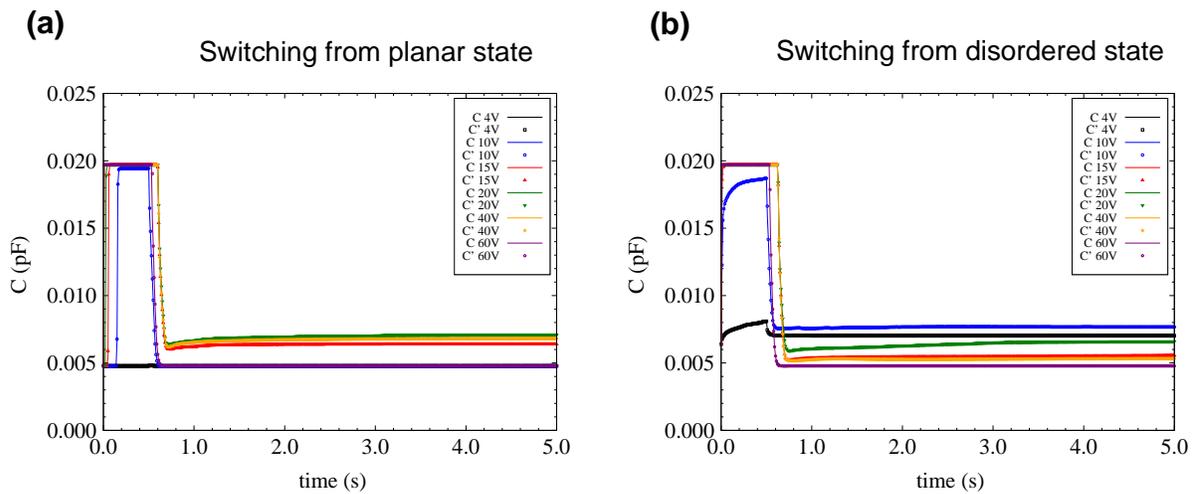

**Figure 6** (Color online) Capacitance as a function of time in the liquid crystal cell during switching at different voltages $V_0$. a) Initial perfect planar state; b) Initial disordered focal conic state. *C* and *C'* correspond with the cell capacitance calculated with the induced charge in the top and bottom substrates respectively. Vertical line shows the time step when the voltage pulse $V_0$ is removed.

*Capacitance analysis:* Figure 6 shows simulation data for capacitance vs. time and applied external electric field. Fig. 6(a) shows data for simulations with a planar initial state, corresponding to the microstructures shown in Fig. 3. Here we observe that the restoration of the helix to perfect planar obtained with $V_0$ = 10V and 60V has faster relaxations than disordered helical textures at intermediate voltages. However, the initial helix unwinding – shown by an initial jump in cell capacitance - is faster when applying a 60V pulse than with a 10V pulse. This suggests that a complete planar-vertical-planar sequence for texture switching could be achieved in shorter time periods with larger voltages. Figure 6(b) shows a similar capacitance measurement when switching from an initially defect-rich focal conic texture, corresponding to microstructures shown in Fig. 5. At 60V, the perfect planar texture relaxes and reaches stability faster than any disordered helical textures observed in the simulation run. This result shows a qualitative agreement with the fast switching behavior observed experimentally in systems with matched elastic constants. The transient planar state was never observed, in agreement with predictions for the case of matched twist and bend elastic constants.

**Different Elastic Constants: Finding the transient planar texture and Helfrich-undulations**

Chiral nematic liquid crystal compounds sandwiched between planar plates have a defect-free ground state with a uniform helical planar texture. If a voltage pulse is applied to a system with similar elastic constants $K_{22} = K_{33}$, the resulting texture will be the perfect planar. If $K_{22} \neq K_{33}$, then the system will form a transient planar texture with pitch $p'=(K_{33}/K_{22})\,p$. Figure 7 shows a simulation study of electric switching between a perfect planar and transient planar by applying a voltage of 10V during 1000 simulation time steps. Simulation box is 10 μm x 10μm x 10 μm. The obtained transient pitch is 7.46 μm while the intrinsic pitch of the chiral compound is 3.33μm. This pitch discrepancy could be given by the anchoring strength contribution to the system free energy and thus director microstructure. Figures 7d and 7e show a slice in yz plane of 7a and 7c respectively.

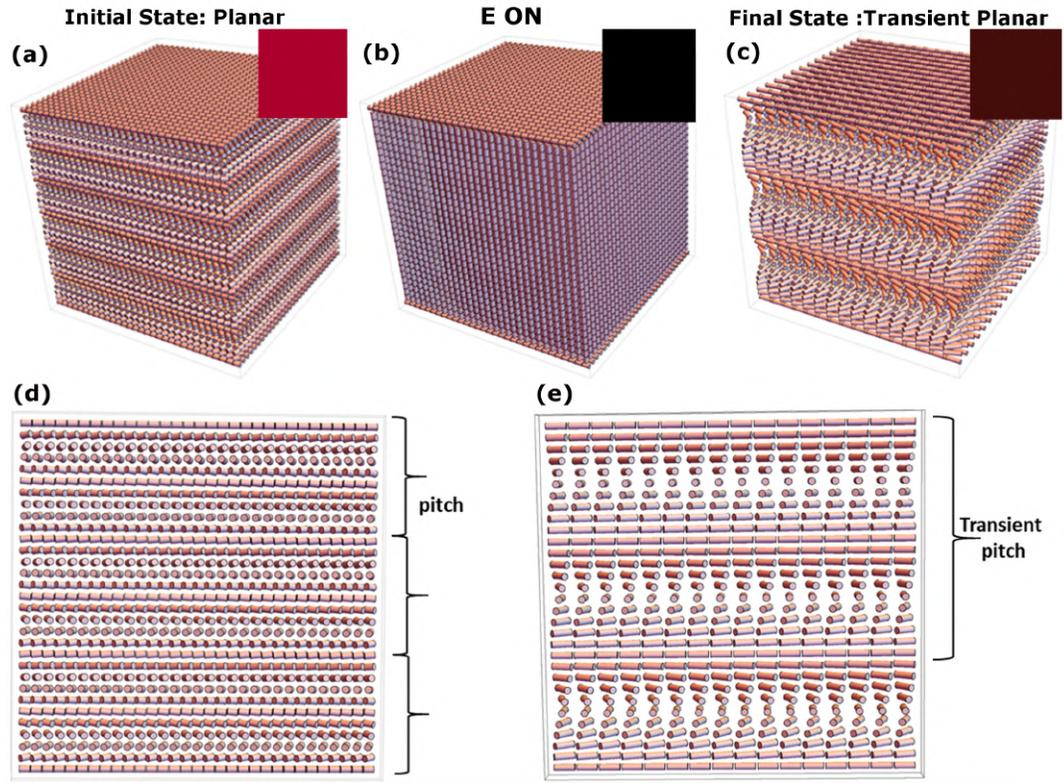

**Figure 7** *Planar to Transient Planar. Relaxation process of a cholesteric liquid crystal starting from planar, applying a 10 v DC pulse and letting it relax to transient planar. The dimensions are 10 μm × 10 μm × 10 μm and the pitch is 3.33 μm and the tran sient pitch is 7.46 μm.*

Similarly, we found a transient planar texture with pitch of ~2.25 μm arising after application of a 40V voltage pulse during 1000 simulation time steps –see Figure 8. The intrinsic pitch of the liquid crystal compound is 1 μm, and the ratio $K_{33}/K_{22} = 2.24$. These simulation results are in agreement with an expected transient pitch of 2.24 μm. We note that the simulation box is 3μm x 3μm x 10μm, and the anchoring surfaces are placed at 0μm and 10μm across the longer axis. This configuration would reduce anchoring effects. Again, cell capacitance falls in a gradual manner.

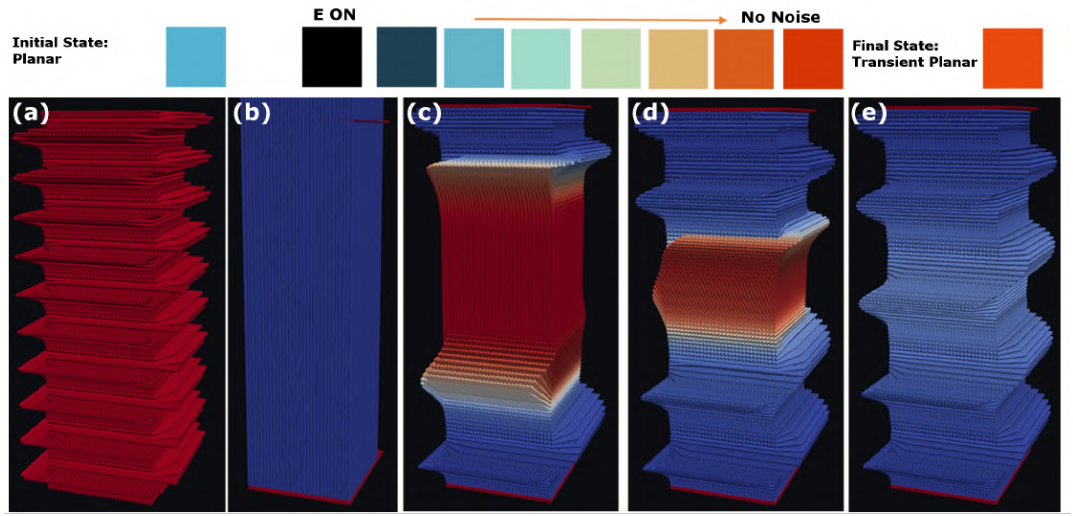

**Figure 8** *Planar to Transient Planar. Relaxation process of a cholesteric liquid crystal starting from planar (a), applying a 40 V DC pulse (b) and letting it relax to transient planar (c-e). The cell dimensions are 3μm × 3μm × 10μm, the intrinsic pitch is 1 μm and the transient planar pitch is 2.24μm.*

Figure 9 shows simulations with an applied thermal noise to the transient planar texture from Figure 7. We found thermal noise distorts the uniform periodicity and allows nucleation of Helfrich-undulations. Figures 9e, 9f and 9g show slices in yz plane of 9a, 9c and 9d respectively. We find that application of thermal noise, before fully relaxing the transient planar, drives formation of the disordered focal conic state. For figures 7 and 9, lattice spacing is 0.333μm. For figure 8, lattice spacing is 0.1μm.

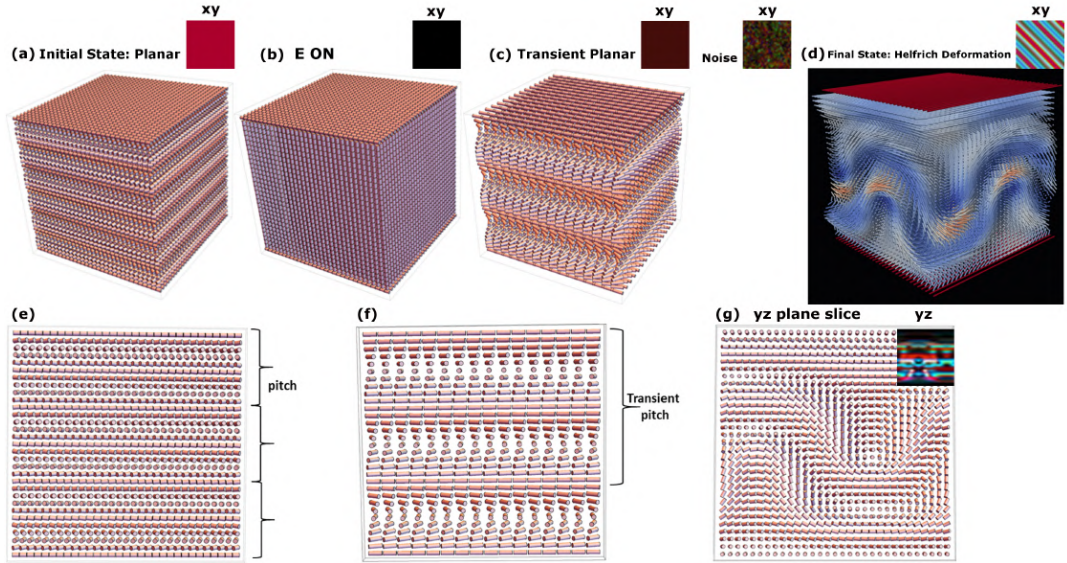

**Figure 9** *Planar to Helfrich deformation. Relaxation process of a cholesteric liquid crystal starting from planar, applying a 10V DC pulse and letting it relax to transient planar and then to Helfrich transition. The dimensions are 10 µm × 10 µm × 10 µm and the pitch is 3.33µm and the transient pitch is 7.46µm.*

This extension to different elastic constants and consequent results show a solid agreement with cholesteric transitions reported in the literature. Model is able to reproduce the arising chiral nematic textures beyond the perfect-planar and focal conic textures, allowing further studies on novel liquid crystal devices. Overall the modeling approaches presented in this work have the capability to set the liquid crystalline bulk in any form; spherical, cubical, oblate, prolate and beyond, opening the door for the study of chiral nematics in different confined geometries with suitable applications in devices as displays, sensors, among others.

**Conclusions**

In this work, we perform a full three-dimensional modeling of texture evolution in chiral nematic liquid crystals by solving for the dynamics of the director field in the presence of a co-evolving electric field. We studied the switching process and electrical response that drives the transition between perfect planar and focal conic textures in samples with equal elastic constants.

Results show agreement with the fast switching behavior at large voltages recently reported in experimental studies in chiral nematics [11]. Furthermore, the resulting focal conic textures in this numerical study present a diverse variety of morphologies. It is noteworthy the texture sequence observed in the focal conic-to-planar study, showing that distortions in the helical axis can have rounded morphologies at small voltages, distorted elongated shapes at intermediate voltages and plain undulations at the larger voltages. A perfect planar texture with fast relaxation is obtained at high voltages, $V_0 = 60V$.

For the case with different elastic constants, chiral nematic texture transitions have been demonstrated. We obtained the expected transient planar texture and were able to find the well-known Helfrich-undulations by the addition of thermal noise in a transient planar texture. In general, the emergence of these textures in our results points that materials with different elastic constants have slower texture switching behavior that those with matched elastic constants.

Overall, we presented a computationally-affordable method to model 3D microstructural evolution in chiral nematic liquid crystals confined between anchoring substrates under a transient applied voltage pulse. This approach allows efficient computation of microstructural evolution in three dimensions and presents capabilities for modeling materials in any confining geometry. These numerical methods show to be a valuable tool for further design, engineering and development of liquid crystalline devices

**Acknowledgments**

We thank S.Y. Lu, and A. Konya for insightful conversations and tests on this simulation approach. We acknowledge J. V. Selinger for his help on the initial development of this model. This work was funded by NSF DMR-1106014, NSF DMR-1409658, the Wright Center of Innovation for Advanced Data Management and Analysis, and by the Ohio Board of Regents. Computing resources provided by the Ohio Supercomputer Center.

**Supplementary Information for:**

**Three-dimensional modeling of chiral nematic texture evolution under electric switching**

Vianney Gimenez-Pinto[*]|| Sajedeh Afghah, and Robin L. B. Selinger[+]

*Advanced Materials and Liquid Crystal Institute, Kent State University, Kent, OH 44242*

||Corresponding author: gimenez-pintov@lincolnu.edu

+Corresponding author: rselinge@kent.edu

*Current Address: Department of Science, Technology and Mathematics, Lincoln University, Jefferson City, MO 65101

*Derivation of the simplified the free energy functional with single elastic constants*

By implementing the single elastic constant approximation $K_{11} = K_{22} = K_{33} = K$, on the style of the Lebwohl-Lasher approach, the system's free energy shown is Eq. 1 can be written as

$$f = K\left(\frac{1}{4}G_1 - q_0 G_4\right) + K G_5.$$

The surface integral involving the $G_5$ term can be neglected because we are modeling the LC bulk. It's straightforward to prove that the quantity $F_1 = \sum_{\langle a,b \rangle}\left(Q_{jk}^{\;b} - Q_{jk}^{\;a}\right)\left(Q_{jk}^{\;b} - Q_{jk}^{\;a}\right)$ is a lattice approximant to $G_1$ — more details of this calculation can be found in our previous work [23] — thus the achiral free energy term follows $F_{achiral} = \frac{K}{4}\int d^3 r\, G_1(\vec{r}) = \frac{K}{4}c\sum_{\langle a,b \rangle}\left(Q_{jk}^{\;b} - Q_{jk}^{\;a}\right)\left(Q_{jk}^{\;b} - Q_{jk}^{\;a}\right)$. In a similar way, using the lattice approximation to $G_4$, $F_4 = \sum_{\langle a,b \rangle}\varepsilon_{jkl}\frac{Q_{jm}^{\;a} + Q_{jm}^{\;b}}{2}(r_l^{\;b} - r_l^{\;a})\left(Q_{km}^{\;b} - Q_{km}^{\;a}\right)$ the chiral free energy goes as

$F_{chiral} = -K q_0\int d^3 r\, G_4(\vec{r}) = -K q_0 c\sum_{\langle a,b \rangle}\varepsilon_{jkl}\left(\frac{Q_{jm}^{\;a} + Q_{jm}^{\;b}}{2}\right)(r_l^{\;b} - r_l^{\;a})\left(Q_{km}^{\;b} - Q_{km}^{\;a}\right)$. Re-expressing the nematic tensor as $Q_{jk} = n_j n_k - \frac{1}{3}\delta_{jk}$, the free energy functional becomes:

$$F = \frac{K}{2}c\sum_{\langle a,b \rangle}\left(1 - \left(\vec{n}^a \cdot \vec{n}^b\right)^2\right) - K q_0 c\sum_{\langle a,b \rangle}\left(\vec{n}^a \cdot \vec{n}^b\right)\left(\vec{n}^a \times \vec{n}^b \cdot \vec{r}^{ab}\right)$$

*Textures transitions from an initial planar state*

Videos 1-7 show the microstructural evolution of the chiral nematic textures when switching from an initial perfect (defect-free) planar state:

Supplementary video 1: $V_0 = 4$ V
Supplementary video 2: $V_0 = 8$ V
Supplementary video 3: $V_0 = 10$ V
Supplementary video 4: $V_0 = 15$ V
Supplementary video 5: $V_0 = 20$ V
Supplementary video 6: $V_0 = 40$ V
Supplementary video 7: $V_0 = 60$ V

*Texture transitions from an initial focal conic state*

Videos 8-16 show the microstructural evolution when switching by applying different $V_0$ voltages to an initial focal conic state:

Supplementary video 8: $V_0 = 4$ V
Supplementary video 9: $V_0 = 6$ V
Supplementary video 10: $V_0 = 8$ V
Supplementary video 11: $V_0 = 10$ V
Supplementary video 12: $V_0 = 12$ V
Supplementary video 13: $V_0 = 15$ V
Supplementary video 14: $V_0 = 20$ V
Supplementary video 15: $V_0 = 40$ V
Supplementary video 16: $V_0 = 60$ V